# Elastic properties of ternary arsenide $SrFe_2As_2$ and quaternary oxyarsenide LaOFeAs as basic phases for new 38-55K superconductors from first principles

I.R. Shein, and A.L. Ivanovskii *

Institute of Solid State Chemistry, Ural Branch of the Russian Academy of Sciences,
620041, Ekaterinburg, Russia

E-mail: shein@ihim.uran.ru and ivanovskii@ihim.uran.ru

**Here we report the first-principle FLAPW-GGA calculations of the elastic properties of two related phases, namely, the ternary arsenide $SrFe_2As_2$ and the quaternary oxyarsenide LaOFeAs as the basic phases for the newly discovered "1111" and "122" 38-52K superconductors. The independent elastic constants ($C_{ij}$), bulk moduli, compressibility, and shear moduli are evaluated and discussed. Additionally, numerical estimates of the elastic parameters of the polycrystalline LaOFeAs and $SrFe_2As_2$ ceramics are performed for the first time.**

Recently, two new groups of FeAs-based high $T_C$ superconductors (SC's) have been discovered [1,2]. For one group (the so-called "1111" type $Ln$OFeAs, where $Ln$ are light rare-earth metals La, Ce….Gd, Tb, Dy), the critical temperature may reach 55K as a result of hole or electron doping [1,3-7]. More recently, ternary oxygen-free arsenides termed as "122" systems, namely, $A$Fe$_2$As$_2$, where $A$ are alkaline earth metals Ca, Sr or Ba, were discovered [2] as parent phases for new hole-doped 20-38K SC's [8-10]. All of these FeAs-based SCs (1) are derivatives of magnetic spin density wave (SDW) systems; (2) adopt layered tetragonal structures, where $[LnO]^{1+}$ layers (or planar sheets of ions $A^{2+}$) are sandwiched between $[FeAs]^{1-}$ layers formed by edge-shared tetrahedra FeAs$_4$ and (3) reveal a two-dimensional electronic structure. It is believed that superconductivity emerges in the [FeAs] layers while the LaO layers (or planar sheets of alkaline earth atoms) provide a charge (hole) reservoir in doping procedures [1-10].

These findings generated a tremendous interest in the scientific community and triggered much activity in research of unconventional superconducting mechanism for these FeAs-based systems. Additionally, some advanced applications for these materials were proposed. For example, the upper critical field $H_{c2}(0)$ for LaO$_{1-x}$F$_x$FeAs can exceed 60 - 65 T [11], making these materials potentially useful in very high magnetic field applications, see also [12]. Others possible applications of these materials may be found in thermoelectric cooling modules in the liquid nitrogen temperature range [13].

Thus, the mechanical properties of FeAs-based SC's are of great importance for their material science in view of future technological applications. Besides, possible correlations between the critical temperature $T_C$ and the mechanical parameters are of interest. For example, it was supposed [14] that high $T_C$'s are associated with low values of the bulk modulus $B$, i.e. with high compressibility $\beta$. Really, many SC's with enhanced critical temperatures (such as YBCO, $MgB_2$, $MgCNi_3$, borocarbides $LnM_2B_2C$, carbide halides of rare earth metals, $Ln_2C_2X_2$ *etc*) are relatively soft: their bulk moduli do not exceed $B \leq 200$ GPa ($\beta \geq 0.005$ 1/GPa) [15-18]. On the other hand, the superconducting transition (to $T_C \sim 11K$) was found for such a hard and incompressible material as boron - doped diamond [19].

The main focus of the present Communication is the *ab initio* prediction of the elastic properties of the ternary arsenide $SrFe_2As_2$ and the quaternary oxyarsenide LaOFeAs (as basic phases for the two above mentioned groups of new 38-55K superconductors) which have not yet been investigated either experimentally or theoretically. As a result, their independent elastic constants ($C_{ij}$), bulk moduli, compressibility, and shear moduli were evaluated and discussed. Additionally, the numerical estimates of the elastic parameters for the polycrystalline $SrFe_2As_2$ and LaOFeAs ceramics were performed for the first time.

Our calculations were carried out by means of the full-potential method with mixed basis APW+lo (LAPW) implemented in the WIEN2k suite of programs [20]. The generalized gradient approximation (GGA) to exchange-correlation potential in the PBE form [21] was used. The calculations were performed with full-lattice optimizations including internal parameters $z_{La}$ and $z_{As}$. The self-consistent calculations were considered to be converged when the difference in the total energy of the crystal did not exceed 0.01 mRy as calculated at consecutive steps, and the nonmagnetic state for $SrFe_2As_2$ and LaOFeAs was treated.

Firstly, the values of six independent elastic constants ($C_{ij}$; namely $C_{11}$, $C_{12}$, $C_{13}$, $C_{33}$, $C_{44}$ and $C_{66}$) for the tetragonal $SrFe_2As_2$ and LaOFeAs phases were evaluated by calculating the stress tensors on different deformations applied to the equilibrium lattice of the tetragonal unit cell, whereupon the dependence between the resulting energy change and the deformation was determined, Table 1. For LaOFeAs, all these elastic constants were positive and satisfied the well-known Born's criteria for tetragonal crystals: $C_{11} > 0$, $C_{33} > 0$, $C_{44} > 0$, $C_{66} > 0$, $(C_{11} - C_{12}) > 0$, $(C_{11} + C_{33} - 2C_{13}) > 0$ and $\{2(C_{11} + C_{12}) + C_{33} + 4C_{13}\} > 0$. However, for $SrFe_2As_2$ our calculations showed $C_{44} \sim 0$. This implies that this phase lies on the border of mechanical stability.

Secondly, the calculated elastic constants allowed us to obtain the macroscopic mechanical parameters for $SrFe_2As_2$ and LaOFeAs phases, namely bulk moduli ($B$) and shear moduli ($G$) – for example, using the Voigt (V) [22] approximation, as:

$$B_V = 1/9\{2(C_{11} + C_{12}) + C_{33} + 4C_{13}\};$$
$$G_V = 1/30(M + 3C_{11} - 3C_{12} + 12C_{44} + 6C_{66});$$

where $C^2 = (C_{11} + C_{12})C_{33} - 2C_{13}^2$ and $M = C_{11} + C_{12} + 2C_{33} - 4C_{13}$. The results obtained are presented in Table 1. From these data we see that for both SrFe$_2$As$_2$ and LaOFeAs phases $B_V > G_V$; this implies that a parameter limiting the mechanical stability of these materials is the shear modulus $G_V$. That is obvious enough due to the layered structure and the sharply anisotropic bonding picture for these phases, as may be clearly seen on the electron density maps, Fig. 1.

Next, as SrFe$_2$As$_2$ and LaOFeAs samples are usually prepared and investigated as polycrystalline ceramics (see [1-12]) in the form of aggregated mixtures of micro-crystallites with a random orientation, it is useful to estimate the corresponding parameters for these polycrystalline materials from the elastic constants of the single crystals. For this purpose we also have calculated monocrystalline bulk moduli ($B$) and shear moduli ($G$) in Reuss approximation (R: $B_R$ and $G_R$) [23], and then we have utilized the Voigt-Reuss-Hill (VRH) approximation. In this approach, according to Hill [24], the Voigt and Reuss averages are limits and the actual effective moduli for polycrystals could be approximated by the arithmetic mean of these two limits. Then, one can calculate the averaged compressibility ($\beta_{VRH} = 1/B_{VRH}$), Young moduli ($Y_{VRH}$), and from $B_{VRH}$, $G_{VRH}$ and $Y_{VRH}$ it is possible to evaluate the Poisson's ratio ($v$). All these parameters are listed in Table 2. Certainly, all these estimations are performed in the limit of zero porosity of SrFe$_2$As$_2$ and LaOFeAs ceramics.

From our results we see that the bulk moduli for SrFe$_2$As$_2$ and LaOFeAs ceramics are rather small (< 100 GPa) and are less than, for example, the bulk moduli for other known superconducting species such as MgB$_2$, MgCNi$_3$, YBCO and YNi$_2$B$_2$C for which $B$ vary from 115 up to 200 GPa [15-18]. Thus, as compared with other superconductors, SrFe$_2$As$_2$ and LaOFeAs are soft materials.

In turn, comparing SrFe$_2$As$_2$ and LaOFeAs, we see that $B_{VRH}$(LaOFeAs) > $B_{VRH}$(SrFe$_2$As$_2$) and $G_{VRH}$(LaOFeAs) > $G_{VRH}$(SrFe$_2$As$_2$). Accordingly, the compressibility of these species changes as $\beta$(SrFe$_2$As$_2$) < $\beta$(LaOFeAs). The Young's modulus $Y_{VRH}$ has also the maximal and minimal values for SrFe$_2$As$_2$ and LaOFeAs ceramics, respectively. A simple explanation of these data follows from the known correlation between the bulk moduli and cell volumes ($B \sim V_o^{-k}$ [25]). In our case, the optimized cell volumes are $V_o$(SrFe$_2$As$_2$) = 183.6 Å$^3$ > $V_o$(LaOFeAs) = 141.2 Å$^3$.

Finally, according to the criterion [26], a material is brittle if the $B/G$ ratio is less than 1.75. In our case, for SrFe$_2$As$_2$ and LaOFeAs these values are 26.83 and 1.74, respectively. This means that LaOFeAs lies on the border of brittle behavior, whereas SrFe$_2$As$_2$ will behave as a very brittle material. Besides, the values of the Poisson ratio ($v$) are minimal for covalent materials ($v = 0.1$) and grow essentially for ionic species [27]. In our case, the values of $v$ for LaOFeAs and SrFe$_2$As$_2$ are about 0.26 and 0.48, respectively, i.e. a considerable ionic contribution in intra-atomic bonding for these phases takes place (see Fig. 1); and this contribution is higher for SrFe$_2$As$_2$.

In conclusion, by performing the first-principles FLAPW-GGA total energy calculations we have predicted the elastic properties for mono- and polycrystalline SrFe$_2$As$_2$ and LaOFeAs as basic phases for the newly discovered 38-55K SCs. Our analysis shows that LaOFeAs is a mechanically stable anisotropic material, whereas SrFe$_2$As$_2$ lies on the border of mechanical stability. The parameter limiting their mechanical stability is the shear modulus. Both phases are relatively soft materials ($B$ < 100 GPa) with high compressibility and will exhibit a brittle behavior.

**Table 1.** Calculated elastic constants ($C_{ij}$, in GPa), bulk moduli ($B$, in GPa) and shear moduli ($G$, in GPa) for tetragonal monocrystalline $SrFe_2As_2$ and LaOFeAs.

| phase / parameters | $SrFe_2As_2$ | LaOFeAs |
|---|---|---|
| $C_{11}$ | 166.1 | 191.9 |
| $C_{12}$ | 30.2 | 55.9 |
| $C_{13}$ | 36.9 | 61.6 |
| $C_{33}$ | 65.0 | 144.8 |
| $C_{44}$ | ~0 | 44.1 |
| $C_{66}$ | 80.5 | 77.9 |
| $B_V$ * | 67.3 | 98.5 |
| $G_V$ * | 32.1 | 56.5 |

* in Voigt approximations

**Table 2**. Calculated values of some elastic parameters for polycrystalline $SrFe_2As_2$ and LaOFeAs ceramics as obtained in the Voigt-Reuss-Hill approximation: bulk moduli ($B_{VRH}$, in GPa), compressibility ($\beta_{VRH}$, in GPa$^{-1}$), shear moduli ($G_{VRH}$, in GPa), Young's moduli ($Y_{VRH}$, in GPa) and Poisson's ratio ($v$).

| oxypnictide | $B_{VRH}$ | $\beta_{VRH}$ | $G_{VRH}$ | $Y_{VRH}$ | $v$ |
|---|---|---|---|---|---|
| $SrFe_2As_2$ | 61.7 | 0.01621 | 2.3 | 6.8 | 0.482 |
| LaOFeAs | 97.9 | 0.01022 | 56.2 | 141.5 | 0.259 |

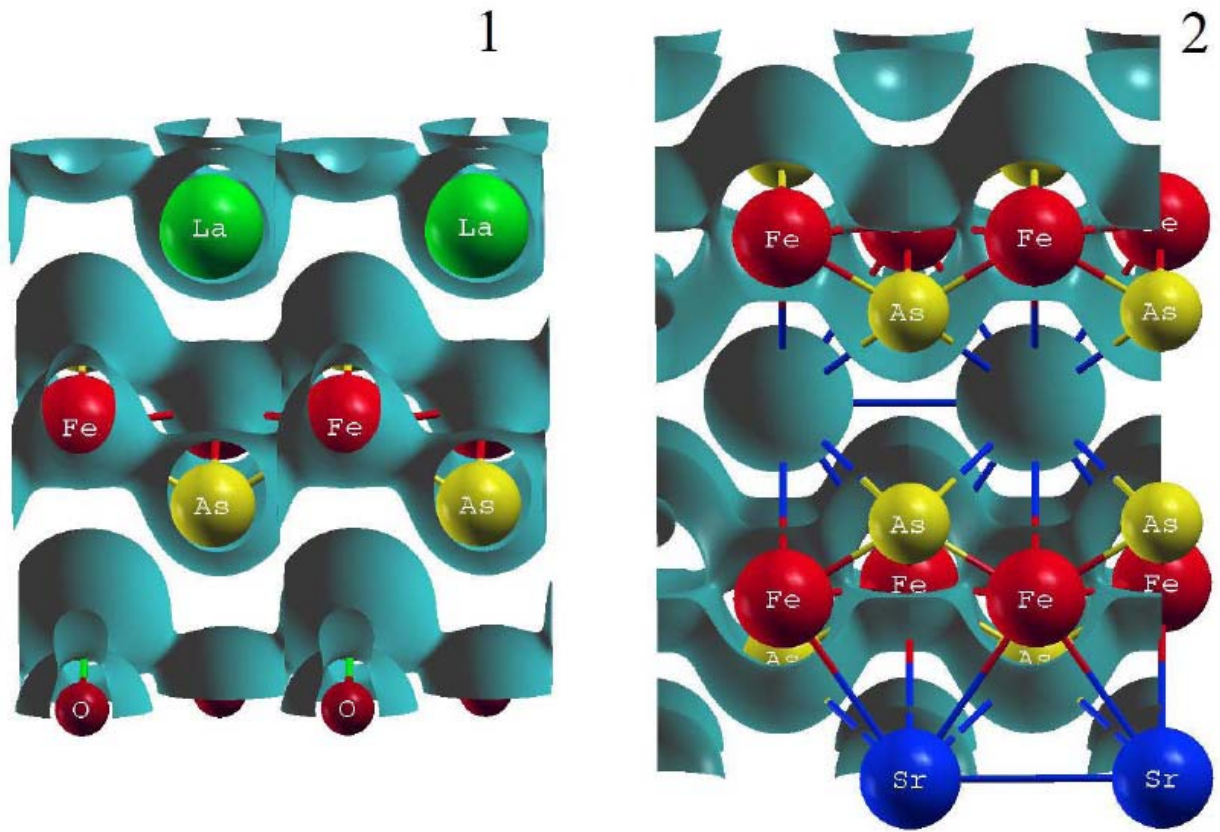

**Figure 1.** (*Color online*) Isoelectronic ($\rho = 0.36$ e/Å$^3$) surface for 1: LaOFeAs and 2: SrFe$_2$As$_2$ according to FLAPW-GGA calculations.